# An electro-optically tunable microring laser monolithically integrated on lithium niobate on insulator

**DiFeng Yin,[1,3] Yuan Zhou,[1,3] Zhaoxiang Liu,[2] Zhe Wang,[1,3] Haisu Zhang,[2] Zhiwei Fang,[2,6] Wei Chu,[2,7] Rongbo Wu,[1,3] Jianhao Zhang,[1,3] Wei Chen,[2] Min Wang,[2] and Ya Cheng[1,2,4,5,8]**

[1]*State Key Laboratory of High Field Laser Physics and CAS Center for Excellence in Ultra-intense Laser Science, Shanghai Institute of Optics and Fine Mechanics (SIOM), Chinese Academy of Sciences (CAS), Shanghai 201800, China*
[2]*XXL—The Extreme Optoelectromechanics Laboratory, School of Physics and Electronic Science, East China Normal University, Shanghai 200241, China*
[3]*Center of Materials Science and Optoelectronics Engineering, University of Chinese Academy of Sciences, Beijing 100049, China*
[4]*Collaborative Innovation Center of Extreme Optics, Shanxi University, Taiyuan 030006, China.*
[5]*Collaborative Innovation Center of Light Manipulations and Applications, Shandong Normal University, Jinan 250358, People's Republic of China*
[6]*zwfang@phy.ecnu.edu.cn*
[7]*chuwei0818@qq.com*
[8]*ya.cheng@siom.ac.cn*

(Dated: March 15, 2021)

**We demonstrate monolithic integration of an electro-optically (EO) tunable microring laser on lithium niobate on insulator (LNOI) platform. The device is fabricated by photolithography assisted chemo-mechanical etching (PLACE), and the pump laser is evanescently coupled into the erbium ($Er^{3+}$) doped LN microring laser using an undoped LN waveguide mounted above the microring. The quality factor of the LN microring resonator is measured as high as $1.54 \times 10^5$ at the wavelength of 1542 nm. Lasing action can be observed at a pump power threshold below 3.5 mW using a 980 nm continuous-wave pump laser. Finally, tuning of the laser wavelength is achieved by varying the electric voltage on the microelectrodes fabricated in the vicinity of microring waveguide, showing an EO coefficient of 0.33 pm/V.**

Monolithic microring lasers have been intensively investigated owing to the potential of producing efficient continuous-wave microlasers over a wide wavelength range using mass-producible microfabrication techniques [1-15]. Currently, tuning of microring lasers have typically been achieved by thermal effect or carrier-injection approach [11, 15]. Electro-optically tunable microring lasers which are capable of providing higher tuning efficiency and speed have not been achieved owing to the weak EO coefficients of current material platforms of high Q microring resonators such as Silicon, $Si_3N_4$ and $Al_2O_3$. This difficulty has recently been overcome by realizing high-Q microring resonators on LNOI [16-20], as LN has a large electro-optic coefficient ($r_{33} = 30.9$ pm/V@$\lambda = 632.8$ nm). Recently, microdisk lasers fabricated on LNOI have been realized, showing low threshold of pump power [21, 22]. However, the microdisk lasers are excited using evanescently coupled pump lasers with tapered fibers, making it difficult to produce monolithically integrated devices. On the other hand, electro-optically tunable microring resonators integrated with electrodes has been demonstrated [23-25]. By producing the same microring resonators on the active $Er^{3+}$ doped LNOI substrate, monolithically integrated microring lasers can be generated without being bothered by the instability inherently related to the tapered fibers. In addition, microelectrodes can be integrated in a straightforward manner with the microring laser to facilitate electro-optic tuning of the laser wavelength. As an indispensable building block for photonic integrated circuit (PIC) applications, the monolithically integrated microring lasers will provide low-threshold, narrow bandwidth, and EO tunable on-chip laser sources for photonics applications.

Commercial (NANOLN, Jinan Jingzheng Electronics Co. Ltd.) $Er^{3+}$ doped and undoped thin film wafers of 600 nm-thick Z-cut lithium niobate on insulator (LNOI) are chosen for device fabrication and bonding process. The concentration of $Er^{3+}$ ions in the LN waveguides is 1 mol%. A schematic diagram of the device fabrication and bonding process is shown in Fig. 1. Firstly, the $Er^{3+}$ doped LN microring resonator was fabricated on Er$^{3+}$ doped thin film LN chip [Fig. 1(a)], meanwhile the undoped LN waveguide was prepared on another undoped LN thin film chip [Fig. 1(b)]. The on-chip undoped LN waveguide and the Er$^{3+}$ doped LN microring resonator were both fabricated by the photolithography assisted chemo-mechanical etching (PLACE) technique, more fabrication details about the PLACE can be find in Ref. 23-25. Secondly, the Er$^{3+}$ doped LN microring resonator thin film chip was mounted on a 6-axis stages and directly attach bonding on the undoped LN waveguide thin film chip [Fig. 1(c)]. The microring should align with the waveguide in a vertical configuration to achieve efficient light coupling [Fig. 1(d)]. Thirdly, to achieve a stable and durable bonding, the ultraviolet (UV) glue is applied to fasten the two chip by irradiation with UV light [Fig. 1(e)]. Finally, monolithic integration of an Er$^{3+}$ doped LN microring resonator with an undoped LN waveguide is achieved [Fig. 1(f) and (g)].

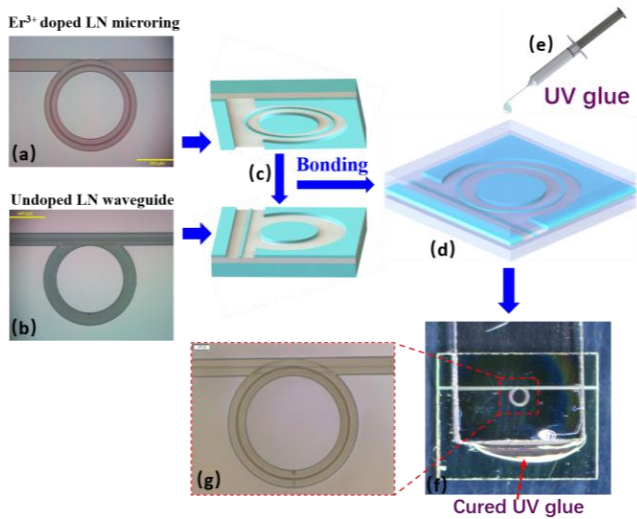

**Fig. 1.** Schematic of the device fabrication and bonding process. (a) The Er$^{3+}$ doped LN microring resonator was fabricated on Er$^{3+}$ doped thin film LN chip. (b) The undoped LN waveguide was prepared on another undoped LN thin film chip. (c)-(d) The Er$^{3+}$ doped LN microring resonator thin film chip directly attach bonding on the undoped LN waveguide thin film chip. (e) The ultraviolet (UV) glue is applied to bond the two chips. (g) Closed-up view of the monolithic integration of an Er$^{3+}$ doped LN microring resonator with an undoped LN waveguide. (f) The photograph of the entire bonding chip.

The experimental setup for characterizing the laser performance of the Er$^{3+}$ doped LN microring resonator was shown in Fig. 2(a). Here, a continuously tunable laser (DLC CTL 1550, TOPTICA Photonics Inc.) was used for characterizing the Q factor of the microring resonator. Alternatively, a diode laser (CM97-1000-76PM, Wuhan Freelink Opto-electronics Co., Ltd.) operated at the wavelength ~980 nm was chosen to pump the Er$^{3+}$-doped LN microring resonator. The polarization states of both the tunable laser and pump laser are adjusted using an in-line fiber polarization controller. The light into and out of the undoped LN waveguide were coupled by a lensed fibers. A photodetector (New focus 1811-FC-AC, Newport Inc.) was directed in the fiber path to measure the Q factor of a resonant mode of the Er$^{3+}$ doped LN microring resonator. The spectrum of the output beam was analyzed by an optical spectrum analyzer (OSA: AQ6370D, YOKOGAWA Inc.). As shown in the inset of Fig. 2, an optical micrograph the strong green upconversion fluorescence in the Er$^{3+}$ doped LN microring resonator pumped by 980 nm laser taken using a COMS camera (DCC3240C, Thorlabs Inc.) mounted onto an optical microscope. Fig. 2(b) shows the transmission spectrum for the wavelength range from 1536 nm to 1543 nm, the free spectral range (FSR) of the 400-μm-diameter Er$^{3+}$ doped LN microring is measured to be 0.805 nm. Fig. 2(c) illustrates that the quality (Q) factor was determined to be 1.54 ×10$^5$ through a Lorentzian fitting of the measured resonance at the wavelength of 1542 nm.

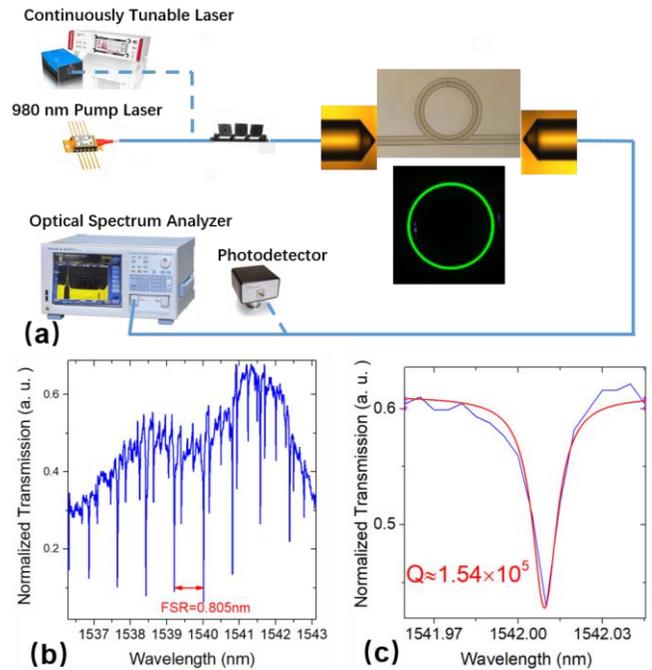

**Fig. 2.** (a) Schematic of the experimental setup for the lasing measurements on the Er$^{3+}$ doped LN microring laser, the inset displays the strong green upconversion fluorescence of the Er$^{3+}$ doped LN microring resonator pumped by 980nm laser. (b) Transmission spectrum of the 400-μm-diamater Er$^{3+}$doped LN microring resonator. (c) The Lorentzian fitting indicating the Q-factors of 1.54×10$^5$ of the microring as measured at 1542 nm wavelength.

To achieve a tunable microresonator laser, a racetrack microring resonator was fabricated on the Er$^{3+}$ doped LN thin film by integrating microelectrodes alongside the two straight arms, the lasing wavelength can be tuned by applying the electric voltage on between the cathode and the two anodes shown in Fig. 3 (a). Fig. 3(a), presenting the top view optical micrograph of the on-chip Er$^{3+}$ doped LN racetrack microring resonator integrated with the Cr electrodes. The diameter of the two half circles of the racetrack microring resonator is 400 μm, and the length of straight arms of the racetrack microring resonator is 500 μm. Therefore, the perimeter of the racetrack microring resonator is ~2.26 mm. The two electrodes on the two sides of the waveguides are separated from each other by ~50 μm. Fig. 3(b) presents the green upconversion fluorescence in the Er$^{3+}$ doped LN racetrack microring resonator pumped by 980 nm laser. Fig. 3(c) shows the measured transmission spectrum in the wavelength range between 1568nm and 1578 nm, the FSR of the racetrack microring resonator was determined to be 0.46 nm. As shown in fig. 3(d), one of the whispering-gallery-modes (WGM) at the resonant wavelength of 1551.82 nm was chosen for the measurement of the Q factor. By fitting with the Lorentz function, Q is determined to be 1.3×10$^5$.

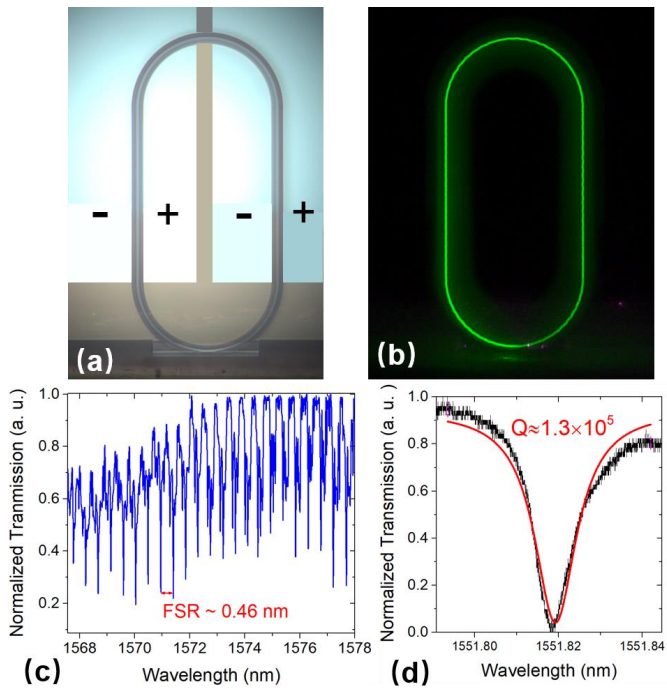

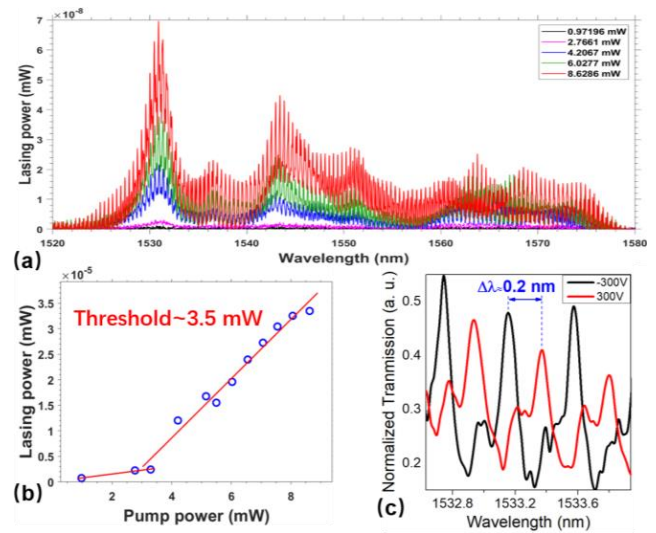

**Fig. 3.** (a) The top view optical micrograph of the on-chip $Er^{3+}$ doped LN racetrack microring resonator integrated with the Cr electrodes. (b) The green upconversion fluorescence in the $Er^{3+}$ doped LN racetrack microring resonator pumped by 980 nm laser. (c) The transmission spectrum of the $Er^{3+}$ doped LN racetrack microring resonator in the wavelength range between 1568nm and 1578 nm. (d) The Lorentzian fitting indicating the Q-factor of $1.3\times10^5$ for the racetrack microring resonator measured at 1551.82 nm wavelength.

The lasing behavior of the $Er^{3+}$ doped LN racetrack microring resonator shows a strong dependence on the pump laser power. Fig. 4(a) shows an emission spectrum collected in the 1520–1580 nm spectral range on the different pump laser power. The lasing emission is highly multimode encompassing 130 narrow-linewidth peaks. And a main emission peak near 1530 nm is recognized, corresponding to the $^4I_{13/2} \rightarrow {}^4I_{15/2}$ transition in $Er^{3+}$. The separation between adjacent peaks in the lasing spectrum is around 0.47 nm, in good agreement with the FSR of LN racetrack microring resonator. As shown in Fig. 4(b), the laser output power integrated over the 1520–1580 nm spectral range is plotted versus the input pump power at 980 nm. A clear lasing threshold is observed around 3 mW, and a linear behavior is observed above the threshold. A direct current (DC) stabilized power source (CE1500002T, Rainworm Co. Ltd.) was used as the voltage generator for Cr electrodes, which provided a variable voltage ranging from 0 to 300 V. A DC probe (ST-20-0.5, GGB Industries Inc.) is used to apply DC voltage on the Cr microelectrodes. As shown in Fig. 4(b), a tuning range of 0.2 nm around the lasing wavelength of 1533 nm can be obtained by varying the electric voltage between -300 V and +300 V, which corresponds to an EO coefficient of 0.33 pm/V. The tuning range is about 42% of the FSR. However, the microelectrodes would break down at a higher voltage.

**Fig. 4.** (a) Spectral evolution of the LN racetrack microring resonator with increasing input pump powers. (b) Laser output power (integrated in the range 1520–1580 nm) as a function of the input pump power, showing a lasing threshold of 3 mW. (c) Lasing line shift by varying the electric voltage between -300 V and +300 V, recorded with a resolution of 0.01nm by OSA.

In summary, we demonstrate a monolithic integration of an electro-optically (EO) tunable microring laser on the thin film lithium niobate (TFLN) platform. The quality factor of the LN microring resonator was measured as high as $1.54 \times 10^5$. Lasing action can be observed with a pump power threshold lower than 3.5 mW when pumping by a 980 nm continuous-wave laser. Besides, the laser wavelength can be tuned in a range of 0.2 nm by varying the electric voltage on the microelectrodes between -300 V and +300 V, which corresponds to an EO coefficient of 0.33 pm/V. The EO tuning efficiency is quite limited in this preliminary investigation because of the employment of Z-cut $Er^{3+}$ doped LNOI wafer which is hard to use the largest EO coefficient ($r_{33}$ for LN) in our current configuration. Future commercial availability of X-cut $Er^{3+}$ doped LNOI wafers will help for producing the EO tunable laser with higher tuning efficiency.

**Funding**.
National Key R&D Program of China (2019YFA0705000), National Natural Science Foundation of China (Grant Nos. 12004116, 11874154, 11734009, 11933005, 11874060, 61991444), Shanghai Municipal Science and Technology Major Project (Grant No.2019SHZDZX01).

**Disclosures.** The authors declare no conflicts of interest

**REFERENCES**

1. D. Liang, M. Fiorentino, T. Okumura, H. Chang, D. T. Spencer, Y. Kuo, A. W. Fang, D. Dai, R. G. Beausoleil, and J. E. Bowers, "Electrically-pumped compact hybrid silicon microring lasers for optical interconnects," Opt. Express **17**, 20355-20364 (2009).
2. J. D. B. Bradley, R. Stoffer, L. Agazzi, F. Ay, K. Wörhoff, and M. Pollnau, "Integrated $Al_2O_3$:$Er^{3+}$ ring lasers on silicon with wide wavelength selectivity," Opt. Lett. **35**, 73-75 (2010).
3. D. Liang, and J. E. Bowers, "Recent progress in lasers on silicon," Nat. Photonics **4**, 511-517(2010).
4. H. Hodaei, M.-A. Miri, M. Heinrich, D. N. Christodoulides, and M. Khajavikhan, "Parity-time-symmetric microring lasers," Science **346**, 975-978 (2014).
5. N. Kobayashi, K. Sato, M. Namiwaka, K. Yamamoto, S. Watanabe, T. Kita, H. Yamada, and H. Yamazaki, "Silicon Photonic Hybrid Ring-

Filter External Cavity Wavelength Tunable Lasers," J. Lightw. Technol. **33**, 1241-1246 (2014).
6. J. D. B. Bradley, E. S. Hosseini, Purnawirman, Z. Su, T. N. Adam, G. Leake, D. Coolbaugh, and M. R. Watts, "Monolithic erbium- and ytterbium-doped microring lasers on silicon chips," Opt. Express **22**, 12226-12237 (2014).
7. Z. Su, N. Li, E. S. Magden, M. Byrd, Purnawirman, T. N. Adam, G. Leake, D. Coolbaugh, J. D. B. Bradley, and M. R. Watts, "Ultra-compact and low-threshold thulium microcavity laser monolithically integrated on silicon," Opt. Lett. **41**, 5708-5711 (2016).
8. N. Li, E. Timurdogan, C. V. Poulton, M. Byrd, E. S. Magden, Z. Su, Purnawirman, G. Leake, D. D. Coolbaugh, D. Vermeulen, and M. R. Watts, "C-band swept wavelength erbium-doped fiber laser with a high-Q tunable interior-ridge silicon microring cavity," Opt. Express **24**, 22741-22748 (2016).
9. D. Liang, X. Huang, G. Kurczveil, M. Fiorentino and R. G. Beausoleil, "Integrated finely tunable microring laser on silicon," Nat. Photonics **10**, 719-722(2016).
10. W. Liu, M. Li, R. S. Guzzon, E. J. Norberg, J. S. Parker, M. Lu, L. A. Coldren and J. Yao, "An integrated parity-time symmetric wavelength-tunable single-mode microring laser," Nat. Commun. **8**, 15389 (2017).
11. N. Li, D. Vermeulen, Z. Su, E. S. Magden, M. Xin, N. Singh, A. Ruocco, J.Notaros, C. V. Poulton, E. Timurdogan, C. Baiocco, and M. R. Watts, "Monolithically integrated erbium-doped tunable laser on a CMOS-compatible silicon photonics platform," Opt. Express **26**, 16200-16211 (2018).
12. H. Zhang, Q. Liao, Y. Wu, Z. Zhang, Q. Gao, P. Liu, M. Li, J. Yao, and H. Fu, "2D Ruddlesden–Popper Perovskites Microring Laser Array," Adv. Mater. **30**, 1706186(2018).
13. C. Zhang, D. Liang, G. Kurczveil, A. Descos, and R. G. Beausoleil, "Hybrid quantum-dot microring laser on silicon," Optica **9**, 1145-1151 (2019).
14. H. Huang, Z. Yu, D. Zhou, S. Li, L. Fu, Y. Wu, C. Gu, Q. Liao, and H. Fu, "Wavelength-Turnable Organic Microring Laser Arrays from Thermally Activated Delayed Fluorescent Emitters," ACS Photonics **6**, 3208-3214(2019).
15. D. Liang, G. Kurczveil, X. Huang, C. Zhang, S. Srinivasan, Z. Huang, M. A. Seyedi, K. Norris, M. Fiorentino, J. E. Bowersb, R. G. Beausoleil, "Heterogeneous silicon light sources for datacom applications," Optical Fiber Technology **44**, 43-52 (2018).
16. M. Zhang, C. Wang, R. Cheng, A. Shams-Ansari, and M. Lončar, "Monolithic ultra-high-Q lithium niobate microring resonator," Optica **4**, 1536 (2017).
17. R. Wu, M. Wang, J. Xu, J. Qi, W. Chu, Z. Fang, J. Zhang, J. Zhou, L. Qiao, Z. Chai, J. Lin, and Y. Cheng, Nanomaterials **8**, 910 (2018).
18. R. Wolf, I. Breunig, H. Zappe, and K. Buse, "Scattering-loss reduction of ridge waveguides by sidewall polishing," Opt. Express **26**, 19815–19820 (2018).
19. J. Lin, F. Bo, Y. Cheng, and J. Xu, "Advances in on-chip photonic devices based on lithium niobate on insulator," Photon. Res. **8**, 1910-1936, (2020).
20. Y. Jia, L. Wang, and F. Chen, "Ion-cut lithium niobate on insulator technology: Recent advances and perspectives," Appl. Phys. Rev. **8**, 011307 (2021).
21. Z. Wang, Z. Fang, Z. Liu, W. Chu, Y. Zhou, J. Zhang, R. Wu, M. Wang, T. Lu, and Y. Cheng, " On-chip tunable microdisk laser fabricated on $Er^{3+}$-doped lithium niobate on insulator," Opt. Lett. **46**,380-383 (2021).
22. Q. Luo, Z. Hao, C. Yang, R. Zhang, D. Zheng, S. Liu, H. Liu, F. Bo, Y. Kong, G. Zhang, and J. Xu, "Microdisk lasers on an erbium-doped lithium-niobite chip," Sci. China Phys. Mech. Astron. **64**,1-5,(2021).
23. A. Guarino, G. Poberaj, D. Rezzonico, R. Degl'Innocenti, and P. Günter, "Electro–optically tunable microring resonators in lithium niobate," Nat. Photonics **1**, 407–410 (2007).
24. C. Wang, M. Zhang, M. Yu, R. Zhu, H. Hu and M. Loncar, "Monolithic lithium niobate photonic circuits for Kerr frequency comb generation and modulation," Nat. Commun. **10**, 978 (2019).
25. Z. Wang, C. Wu, Z. Fang, M. Wang, J. Lin, R. Wu, J. Zhang, J. Yu, M. Wu, W. Chu, T. Lu, G. Chen, and Y. Cheng, "High-quality-factor optical microresonators fabricated on lithium niobite thin film with an electro-optical tuning range spanning over one free spectral range," Preprints 2020070685 (2020).
26. R. Wu, J. Zhang, N. Yao, W. Fang, L. Qiao, Z. Chai, Ji. Lin, and Y. Cheng, "Lithium niobate micro-disk resonators of quality factors above $10^7$," Opt. Lett. **43**, 4116 (2018).
27. M. Wang, R, Wu, J. Lin, J, Zhang, Z. Fang, Z, Chai and Y. Cheng, "Chemo-mechanical polish lithography: A pathway to low loss large-scale photonic integration on lithium niobate on insulator," Quantum Engineering **1**, e9 (2019).
28. Z. Fang, S. Haque, S. Farajollahi, H. Luo, J. Lin, R. Wu, J. Zhang, Z. Wang, M. Wang, Y. Cheng, and T. Lu, " Polygon Coherent Modes in a Weakly Perturbed Whispering Gallery Microresonator for Efficient Second Harmonic, Optomechanical, and Frequency Comb Generations," Phys. Rev. Lett. **125**, 173901 (2020).